\documentclass[sigconf]{acmart}
\AtBeginDocument{%
  }

\setcopyright{acmlicensed}
\copyrightyear{2018}
\acmYear{2018}
\acmDOI{XXXXXXX.XXXXXXX}
\acmConference[Conference acronym 'XX]{Make sure to enter the correct
  conference title from your rights confirmation email}{June 03--05,
  2018}{Woodstock, NY}
\acmISBN{978-1-4503-XXXX-X/2018/06}

\usepackage{graphicx}
\usepackage{subcaption}
\usepackage{multirow}
\usepackage{adjustbox}

\usepackage{tabularx}
\usepackage[table]{xcolor} 
\usepackage{array}
\usepackage{ragged2e} 
\usepackage{enumitem}

\usepackage{booktabs}
\usepackage{array}
\usepackage{colortbl}
\usepackage{arydshln}

\usepackage[utf8]{inputenc}
\usepackage[T1]{fontenc}
\usepackage[linesnumbered,ruled,vlined]{algorithm2e}
\usepackage{amsmath}

\begin{document}

\title{LLM4AIGQ: LLM-based AI Guidance Query Generation Framework for Multi Interest Mining}

\author{Xiangchen Pan}
\affiliation{%
  \institution{Huazhong University of Science and Technology, Alibaba Group}
  \city{Wuhan}
  \country{China}}
\email{pxcstart666@gmail.com}

\author{Jiayi Xu}
\affiliation{%
  \institution{Alibaba Group}
  \city{Hangzhou}
  \country{China}}
\email{youyi.xjy@alibaba-inc.com}

\author{Jing Wang}
\affiliation{%
  \institution{Alibaba Group}
  \city{Hangzhou}
  \country{China}}
\email{jing.wangj1@taobao.com}

\author{Xing Fang}
\authornote{Corresponding author.}
\affiliation{%
  \institution{Nankai University, Alibaba Group}
  \city{Hangzhou}
  \country{China}
}
\email{fangxing.fx@taobao.com}

\author{Lingyun Zhu}
\affiliation{%
  \institution{Alibaba Group}
  \city{Hangzhou}
  \country{China}}
\email{zly234561@taobao.com}

\begin{abstract}
Guidance queries stimulate user consumption by extracting preferences to provide search queries with guidance value, playing a crucial role in the e-commerce field. Traditional AI-generated queries(AIGQ) generation primarily relies on a two-stage "Query-to-Ai-Generated-Query"(Q2AIGQ) association paradigm, first recalling user primary search queries from user profiles, historical behavior sequences, item-side information, and the current query through multi-path retrieval, then generalizing AIGQ via rule-based methods. This approach suffers from semantic drift due to information cascade loss; additionally, primary search query derivation heavily depends on "user-item" co-occurrence relationships, lacking exploration of user multi-interests, resulting in guidance queries with low value and mismatched purchase intent. To address the expressive limitations of traditional co-occurrence-based retrieval, we propose LLM4AIGQ, an LLM-based solution for generating AI guidance queries tailored to users’ multi-interests. This approach segments user interests by integrating user profiles and historical interaction sequences, infers specific consumption intents for each sub-interest, and subsequently generates corresponding AIGQ. In terms of model training, we employ a post-training pipeline comprising Supervised Fine-Tuning (SFT), Reinforcement Learning (RL), and Direct Preference Optimization (DPO) to enhance the model’s capability in generating AIGQ. We also introduce a multi-level reward design to satisfy the requirements of multi-objective optimization and long-chain reasoning in practical applications. Regarding deployment, we adopt an nearline-generation and online-read architecture to meet latency constraints. Extensive experimental analyses demonstrate that our model achieves robust performance in both offline evaluations and online A/B tests.
\end{abstract}


\maketitle

\section{Introduction}
Shopping agents provide product recommendations and decision support through natural-language interactions, yet users may struggle to articulate their needs at the outset. To lower this interaction barrier, e-commerce platforms proactively present AI-generated queries (AIGQs) as guidance queries in scenarios such as homepage recommendation and AI-powered search. Once selected, an AIGQ serves as an initial request to the shopping agent. It should therefore reflect user preferences and convey actionable shopping intent, thereby improving user engagement and conversion.Existing methods commonly follow a two-stage Q2AIGQ paradigm: they first infer the user’s primary search query~\cite{Li_Li_Wu_Zhou_Wang_2019,he2022query,si2023search} through multiple reasoning paths based on the user profile and interaction history, item-side information, and the current query, and then, transform them into AIGQs through rule-based generalization. This intermediate-query bottleneck causes information loss and semantic drift. Moreover, anchor-query inference is often dominated by isolated search behaviors and user-item co-occurrence signals, limiting its ability to model users' diverse interests and personalized intent. Consequently, the resulting AIGQs may provide limited guidance and poorly align with users' actual shopping intent.

\begin{figure}[t]
    \centering
    \includegraphics[width=1\linewidth]{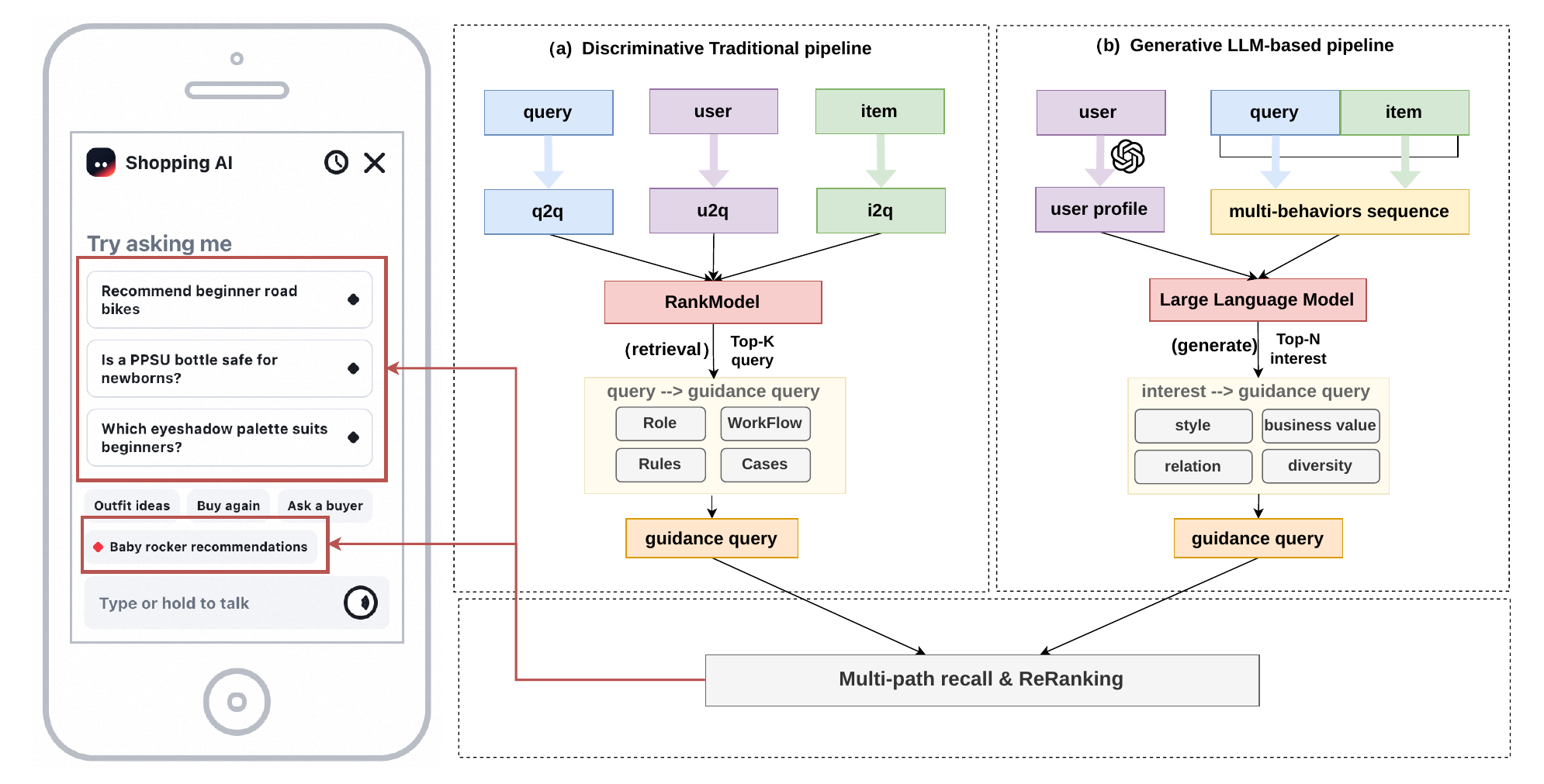}
    \caption{Comparison between Generative Pipeline and Traditional Discriminative Pipeline in the Development of Guidance queries}
    \label{intro}
\end{figure}

Large language models (LLMs) have been widely applied in recommendation systems~\cite{liu2024llm, pan2026mmp} due to their strong semantic understanding and text generation capabilities. For example, XRec~\cite{ma2024xrec} maps collaborative features into the semantic space of LLMs for explainable recommendation; ReFICR~\cite{yang2024unleashing} enables retrieval-augmented LLMs to perform conversational recommendation through instruction tuning; and Rec-R1~\cite{lin2025rec} leverages reward-driven reinforcement learning to help LLMs better understand users’ implicit intents and generate outputs aligned with user needs. As illustrated in Fig. ~\ref{intro}, compared with traditional discriminative multi-channel retrieval methods, LLM-based generative query generation can better capture personalized user intent and avoid semantic drift caused by information cascade loss. Motivated by this, we explore the use of LLMs to infer users’ shopping guidance intent from user profiles and historical behavior sequences, thereby generating AIGQ with higher commercial value and stronger alignment with user purchase intentions.

However, given the vast diversity of items and user groups, efficiently generating accurate and personalized AI-powered guidance queries remains challenging. Specifically, three key challenges need to be addressed: (1) \textbf{Multi-interest modeling under noisy interactions.} Users typically exhibit diverse consumption interests, while their historical interaction sequences inevitably contain noise caused by accidental interactions~\cite{wang2025unleashing, pan2026joint, xv2024improving}, exposure bias, and so on. Therefore, it is crucial to effectively mitigate the impact of such noise and accurately identify users’ primary interests. (2) \textbf{Effective supervision without explicit ground truth.} AI-generated guidance queries do not have strictly defined ground-truth answers, making it challenging to reliably assess their quality and construct effective supervision signals that guide the model to optimize in a reasonable direction. (3) \textbf{Efficient inference for online serving.} Online dynamic inference requires high responsiveness and low latency. Therefore, substantially reducing LLM inference latency while preserving generation quality is essential for satisfying the real-time requirements of practical e-commerce systems.

To address the first challenge, we design task-specific instructions to enable the LLM to learn how to disentangle users’ diverse interests and generate AIGQ for each identified sub-interest. Specifically, user profiles mainly reflect relatively stable long-term interests, whereas behavioral sequences capture short-term interests that are often more dynamic and heterogeneous~\cite{wang2022target, wu2025gemirec, yi2026dualgr}. We first employ a state-of-the-art LLM as the teacher model to decompose each user’s historical interaction sequence. Through sequence merging and pruning, we construct multiple subsequences, each representing a coherent and distinct interest, and use them as supervision labels. These interest-specific subsequences, together with the corresponding user profiles, are then used as training data to guide the LLM in discovering and modeling multiple user interests.

To address the second challenge, we build upon supervised fine-tuning (SFT) and further conduct reinforcement learning (RL) with a multi-objective reward design. Specifically, we first employ a state-of-the-art LLm as teacher model to partition user interests and generate a guidance query for each identified sub-interest, which serves as the reference output. During the SFT stage, the LLM is provided only with single-interest subsequences, with the primary objective of learning the desired generation style and patterns of guidance queries. During the RL stage, in contrast, the instruction data consist of user behavior sequences involving multiple mixed interests, requiring the model to autonomously explore and infer the underlying interest partitioning. The task instructions explicitly specify the expected reasoning process and output format. We further design reward functions from multiple levels for the reasoning (think) and final answer components separately, thereby jointly supervising the model’s intent understanding and text generation capabilities.

To address the third challenge, we adopt two complementary strategies to improve inference efficiency. First, we employ a nearline generation and online retrieval deployment paradigm, where guidance queries are generated in advance through nearline inference and directly retrieved during online serving, thereby substantially reducing the latency requirements imposed on LLM inference. Second, we further apply Direct Preference Optimization (DPO)~\cite{Rafailov_Sharma_Mitchell_Ermon_Manning_Finn} to enable the model to preserve the reasoning effectiveness of the think mode under a non-think inference mode. By eliminating explicit reasoning traces and substantially reducing the number of generated tokens, this strategy accelerates LLM inference while maintaining the quality of generated guidance queries.

In summary, we propose LLM4AIGQ, an LLM-based framework for multi-interest-aware AI guidance query generation, with a three-stage SFT–RL–DPO training pipeline. The main contributions are summarized as follows.

\begin{itemize}[leftmargin=*] 
    \item We propose LLM4AIGQ, an LLM-based framework for multi-interest-aware AI guidance query generation, addressing cascading information loss and insufficient personalized intent modeling in traditional pipelines.
    \item We adopt a three-stage SFT–RL–DPO training pipeline. The SFT stage focuses on learning the style of AIGQ, while the RL stage uses multi-objective rewards to supervise interest disentanglement and high-quality query generation. The DPO stage further optimizes the model’s reasoning expression in non-think mode through preference optimization.
    \item For deployment, we adopt a nearline generation and online retrieval architecture. User guidance query pools are updated offline daily, while nearline LLM inference dynamically generates queries based on recent interactions. The system satisfies strict online latency requirements and achieves strong performance in both offline and online evaluations.
\end{itemize}




\section{Related Work}

\subsection{Query Recommendation}
Traditional query recommendation mainly focuses on query–item relevance matching. Methods such as QEM~\cite{ai2019zero} and DREM~\cite{ai2019explainable} estimate user interests based on semantic similarity between queries and items. To better capture personalized intent, subsequent studies incorporate user profiles and historical search behaviors. HEM~\cite{ai2017learning} models long-term preferences from reviews via user embeddings, while AEM uses attention to aggregate historical interactions conditioned on the current query. ZAM~\cite{ai2019zero} extends AEM with a zero-attention vector to adaptively control personalization, and TEM employs a Transformer encoder to capture sequential dependencies in search histories. Some studies further explore multi-interest modeling. CAMI~\cite{liu2022category} disentangles multiple interests using knowledge graph embeddings, while GraphSRRL~\cite{liu2020structural} exploits structural patterns in user–query–item graphs to enhance interest representations. Recently, given the complementarity between search and recommendation behaviors, methods such as USER~\cite{yao2021user}, UniSAR~\cite{shi2024unisar}, and MIJSR~\cite{pan2026multi} jointly model both scenarios for query recommendation. Unlike conventional query recommendation, our task not only identifies suitable items that match user interests as search entities, but also improves the shopping guidance value to stimulate consumption. We therefore incorporate multi-behavior data and model both long- and short-term interests to capture personalized intent from multiple perspectives.

\subsection{LLM-based Recommendation}
With strong semantic understanding and text generation capabilities, LLMs have attracted increasing attention in recommender systems~\cite{pan2026curr,xiang2026user,zheng2025deeprec}. Early studies mainly used LLMs to enhance textual representations and capture semantic correlations for better recommendation. For example, LLMRec~\cite{wei2024llmrec} leverages LLMs to augment user–item interactions and textual information for graph-based recommendation. Other studies align LLMs with downstream recommendation tasks through instruction tuning, such as TALLRec~\cite{bao2023tallrec}. With advances in chain-of-thought reasoning and reinforcement learning, recent studies have explored various combinations of RL and SFT to further improve reasoning capabilities. In recommendation, methods such as Rec-R1~\cite{lin2025rec}, Reason4Rec~\cite{fang2025reason4rec}, and OneReason~\cite{team2026onereason} employ RL algorithms such as PPO and GRPO~\cite{Shao_Wang_Zhu_Xu_Song_Zhang_Li_Wu_Guo} with tailored reward models to enhance preference understanding and factual reasoning. Our work adopts an SFT–RL–DPO post-training pipeline to improve guidance query generation, with multi-dimensional rewards to support multi-objective optimization and long-chain reasoning.
\section{Preliminary}
This work uses LLM to infer AIGQ based on users' multiple interests. The following introduces the data composition and task objectives.

\textbf{Data Composition:} The input for LLM-based generation of AI guidance queries mainly consists of two components: the task instruction $R$ and the user context $C_u$. The user context is further composed of the user profile $P_u$ and the user historical sequence $S_u$. The user profile is summarized and extracted by the Qwen model based on the user’s consumption patterns and basic personal information. The user historical sequence $S_u$ primarily contains the user’s recent behaviors, where each behavior $B_u$ consists of a timestamp $t$, behavior type $a$, and behavior content $c$. The behavior types $a$ include like, favorite, add-to-cart, purchase, and search. For search behaviors, the behavior content $c$ corresponds to the user’s query terms; for all other behaviors, it corresponds to the associated item titles.

\textbf{Task Objective:} The LLM is required to identify user interests $I_u$ based on the provided user context $C_u$ and generate AIGQ for each sub-interest. The summarized interests should closely align with the user’s consumption patterns, while the generated guidance queries should effectively address user purchase pain points, provide shopping-guidance value, and remain concise and compliant.

\begin{figure}[t]
    \centering
    \includegraphics[width=1\linewidth]{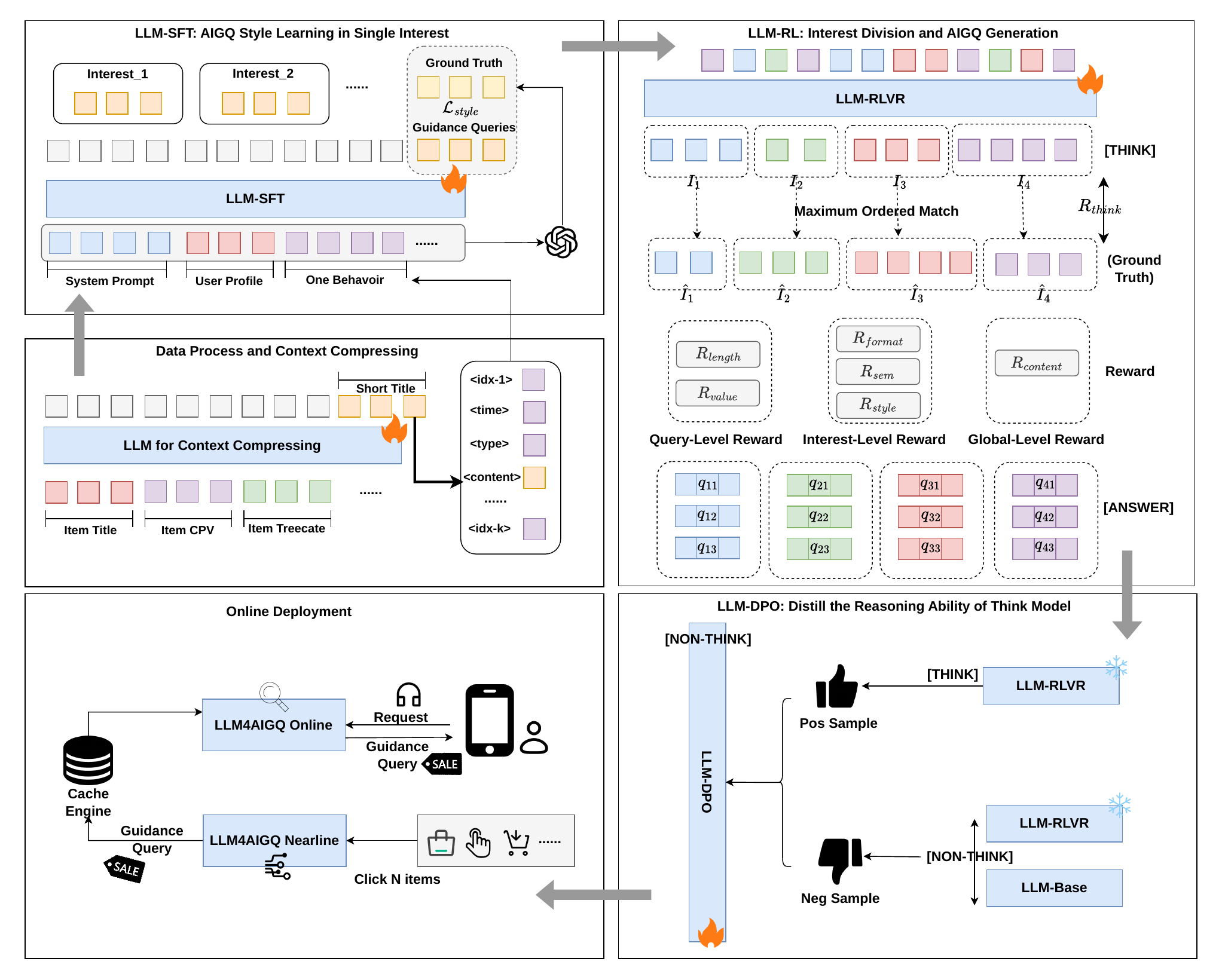}
    \caption{The overview framework of LLM4AIGQ. In the data preprocessing stage, an LLM is used to generate item short titles from item-side information, enabling compressed extraction of behavioral content.The SFT stage is conducted under single-interest scenarios to guide the learning of AIGQ styles. The RL stage adopts multi-level reward functions to supervise interest disentanglement and query generation. DPO further learns the reasoning expression of think mode through preference optimization. For deployment, we adopt a nearline generation and online retrieval architecture.}
    \label{overview}
\end{figure}

\section{Methodology}
In this section, we introduce the LLM4AIGQ framework in detail, as illustrated in Fig. ~\ref{overview}. The framework adopts a three-stage training paradigm: SFT → RL → DPO. The SFT stage enables the LLM to learn the style of AIGQ through supervised learning. The RL stage trains the LLM to perform interest segmentation and infer consumption intent under complex scenarios. The DPO stage distills the reasoning capability of the LLM in think mode, enabling high-quality reasoning outcomes in non-think mode. For model deployment, we adopt a nearline generation and online retrieval architecture. The following sections provide detailed descriptions of each component.

\subsection{Data Process and Context Compression}
Before model training, considering the large volume of users’ historical consumption data, we truncate the historical interaction sequence to reduce unnecessary computational overhead. Specifically, only the most recent 50 interaction behaviors are retained to capture users’ short-term interests. For each interaction behavior $B_u$, we represent it in the format "<Index number $idx$><Time $t$><Behavior type $a$>Behavior content $c$", with behaviors separated by "|". The time information $t$ denotes the interval relative to the current time. Depending on the behavior type $a$, the behavior content $c$ corresponds to either the associated item title or the user’s search query.

The original item titles lack standardized naming conventions and have poor readability; And many titles have a lot of information redundancy in order to pursue search exposure. In order to simplify the expression of behavior sequences, we attempted to use the Qwen3-30B-A3B model to summarize a short item title from the original item title $i_d$, cpv information $i_c$, and brand category information $i_b$, removing redundant information such as marketing keywords and flagship stores.

In order to make the item short titles inferred by LLM concise and effective, we use the item short titles inferred by the larger-capacity model (deepseek-V4 Pro) as the standard answer to fine tune the item short title task for LLM. The optimization objective can be seen in Equ. ~\ref{short_title}:

\begin{equation}
\label{short_title}
\begin{aligned}
\mathcal{L}_{short} = -E_{(x,z)\sim \mathcal{D}}[\sum_{t=1}^TlogP(z_t|z_{<t},x;\theta)]
\end{aligned}
\end{equation}

Where $x=[i_d,i_c,i_b]$ represents the model input, which is item related information, $\theta$ is the model parameter, and $T$ is the length of the generated short title. Here are the instructions for the short title task.

\fbox{
\parbox{0.4\textwidth}{
\textbf{System Prompt:} Please extract item short titles based on item information and map the item to short query text that reflects core selling points and user concerns for representation.

\textbf{User Prompt:} 
\{ \\
"title": "[Hot Selling] Haier Refrigerator Double Door Air Cooled Frost Free Household Energy saving Large Capacity", \\
"cpv": {"Brand": "Haier", "Style": "Double Door", "Cooling Method": "Air Cooled"}, \\
"treecate\_full\_cath": "Household; Refrigerator; Double Door Ice Box", \\
"treecate\_name": "Double Door Refrigerator" \\
\}

\textbf{Output:} \{Predicted short titles.\}
}
}

\subsection{SFT Training: AIGQ Style Learning in Single Interest}
The aim of SFT stage is to help LLM understand the definition of AIGQ and what constitutes good AIGQ. To this end, we will indicate in the task instructions the possible application scenarios of AIGQ (single item consultation, comparative decision-making, category selection, etc.), and require the model to output AIGQ that have guidance value while ensuring content compliance and concise expression. However, through manual testing, it was found that the AIGQ generated directly using the Qwen3-30B-A3B~\cite{yang2025qwen3} model have problems such as low guidance value and poor expression diversity. In contrast, the AIGQ generated by the larger-capacity model (deepseek-V4 Pro) are more in line with practical application scenarios in terms of expression and have greater guidance value. Therefore, we use the inference results of the larger-capacity model as the standard answer, and enhance the model's understanding of AIGQ through instruction fine-tuning.

The historical behavior sequence of users is usually complex and diverse, and may contain multiple user interests. After practical testing, it has been found that directly using historical behavior data of composite interests to fine tune the model can easily lead to the problem of autonomous extrapolation during inference, and is not friendly to low-frequency trigger words. The generated multiple candidate AIGQ often focus on a single main interest.

To address the above issue, we first employ a state-of-the-art LLM to perform interest segmentation on the user historical sequence $S_u$. For each sub-interest $I_k$, the corresponding subsequence $S_k$ is extracted in chronological order, and the associated guidance queries is inferred accordingly. This design encourages the model to focus on learning the style of guidance queries generation rather than selecting among multiple interests. The objective function for this stage is defined as follows:

\begin{equation}
\begin{aligned}
\mathcal{L}_{style} = -E_{(P_u,S_u,z)\sim \mathcal{D}}[\sum_{k=1}^K\sum_{t=1}^TlogP(y_t|y_{<t},P_u,S_k;\theta)]
\end{aligned}
\end{equation}

where $P_u$ denotes the user profile, $\theta$ represents the model parameters (Qwen3-30B-A3B in our implementation), and $T$ denotes the length of the generated AIGQ.

\subsection{RL Training: Interest Division and AIGQ Generation}
This stage aims to guide the LLM to autonomously perform interest segmentation and generate corresponding AIGQ for each sub-interest through multi-dimensional reward design. To better supervise model training, the LLM is instructed to conduct step-by-step reasoning according to the prompt instructions before producing the final answer. An example prompt is shown below:

\fbox{
\parbox{0.4\textwidth}{
\textbf{Think step by step}\\
\textbf{Step 1:} Summarize the user's multiple interests and divide them into sub sequences based on the mixed sequence of multiple interests.\\
\textbf{Step 2:} Infer consumption intentions for each subsequence  and make predictions for the next item.\\
\textbf{Step 3:} Generate 3 corresponding AI guidance queries for each subsequence and provide normalized confidence scores.\\
\textbf{Answer}\\
\{ \\
    "interest\_tag": [query1, query2, query3], \\
    "interest\_tag": [query1, query2, query3], \\
    ... \\
\}
}
}

For Step 1, the model is required to segment the historical sequence into at most three interest subsequences based on temporal proximity of user behaviors, semantic relevance, and profile preferences. For each subsequence $S_k$, the corresponding interaction behaviors $B_k=[B_{k_1},...,B_{k_m}]$ are identified, and an interest label $I_k$ is summarized.

For Step 2, the model is required to infer the user’s current consumption intent $M_k$ based on the interest label $I_k$, the item characteristics reflected in $B_k$, and the user profile $P_u$, while also predicting the next potentially preferred item $i_k$.

For Step 3, the model is required to generate three guidance queries according to the inferred interest $I_k$ and the predicted item $i_k$, and assign a confidence score $c$ to each guidance query.

After completing the above reasoning process, the model outputs the inferred interest labels and their corresponding guidance queries in JSON format as the final answer.

After imposing strict constraints on the model’s reasoning process, we further design multiple reward functions to improve training effectiveness. Considering that multi-objective optimization may introduce issues such as objective conflicts and diluted training signals, we adopt a hierarchical reward design. Specifically, rewards are defined at four granularities: think-level, query-level, interest-level, and global-level, enabling supervision of the model’s reasoning accuracy from multiple perspectives.

\subsubsection{Think-Level Reward}
The think-level reward is designed to supervise the rationality of interest segmentation and the accuracy of interest mining. We first employ a high-parameter model (DeepSeek-V4-Pro) to generate reference reasoning processes and standard outputs under the same prompt settings. Assume that the reference interest subsequences are denoted as $\hat{S}_u=[\hat{S}_1,...,\hat{S}_m]$, with corresponding interest labels $\hat{I}_u=[\hat{I}_1,...,\hat{I}_m]$. The model-generated subsequences are denoted as $S_u=[S_1,...,S_n]$, with corresponding interest labels $I_u=[I_1,...,I_n]$. Since the number of inferred interests may differ between the reference output and the model output, and considering that LLMs generally place primary interests earlier and secondary interests later during reasoning, we adopt an optimal ordered matching algorithm to obtain the most reasonable alignment strategy $M$. The implementation details of the matching algorithm are provided in Alg. ~\ref{alg:max_ordered_match}. The matching score between $\hat{I}_i$ and $\hat{I}_j$ is mainly determined by the overlap between their corresponding interest subsequences $\hat{S}_i$ and $S_j$. Moreover, since matching primary interests is more important during sub-interest alignment, sequence length is introduced as a penalty factor to refine the matching rule. The matching score is computed as follows:

\begin{equation}
\begin{aligned}
score(\hat{I}_i,I_j)=\frac{2*(|\hat{S}_i|\cap|S_j|)}{|\hat{S}_i|+|S_j|} * log(|\hat{S}_i|+|S_j|)
\end{aligned}
\end{equation}

\begin{algorithm}
\caption{Optimal Ordered Matching Algorithm}
\label{alg:max_ordered_match}
\KwIn{
    Sequence $\hat{S}_u$ of length $m$, 
    Sequence $S_u$ of length $n$, 
    Score function $\text{score}(a, b)$
}
\KwOut{
    Maximum score value, DP table, and Parent pointers for backtracking
}

$m \leftarrow \text{length}(\hat{S}_u)$\;
$n \leftarrow \text{length}(S_u)$\;

Initialize $DP[0..m][0..n]$ with zeros\;
Initialize $Parent[0..m][0..n]$ with null\;


\For{$i \leftarrow 1$ \KwTo $m$}{
    \For{$j \leftarrow 1$ \KwTo $n$}{
        $best \leftarrow DP[i-1][j]$\\
        $Parent[i][j] \leftarrow (i-1, j, \text{'skip $\hat{S}_u$'})$\;

        \If{$DP[i][j-1] > best$}{
            $best \leftarrow DP[i][j-1]$\\
            $Parent[i][j] \leftarrow (i, j-1, \text{'skip $S_u$'})$\;
        }

        $matchVal \leftarrow DP[i-1][j-1] + \text{score}(\hat{S}_u[i-1], S_u[j-1])$\;
        \If{$matchVal > best$}{
            $best \leftarrow matchVal$\\
            $Parent[i][j] \leftarrow (i-1, j-1, \text{'match'})$\;
        }

        $DP[i][j] \leftarrow best$\;
    }
}

\Return $(DP[m][n], DP, Parent)$\;
\end{algorithm}

After obtaining the matching set $M$, we compute the interest accuracy reward $R_{match}$ using an action-logic matching strategy. For each matched pair $(a,b) \in M$, the action reward $R_{action}$ is calculated based on the number of matched behaviors between the corresponding subsequences $\hat{S}_a$ and $S_b$, while the logic reward $R_{logic}$ is computed according to the semantic similarity between the corresponding interest labels $\hat{I}_a$ and $I_b$. The overall formulation of $R_{match}$ is defined as follows:

\begin{equation}
\begin{aligned}
R_{match} &= \frac{1}{|M|}\sum_{(a,b)\in M}R_{a,b}=\frac{1}{|M|}\sum_{(a,b)\in M}R_{action}^{(a,b)}*R_{logic}^{(a,b)} \\
R_{(a,b)}^{action} &= \frac{2*(|\hat{S}_a|\cap|S_b|)}{|\hat{S}_a|+|S_b|}, R_{(a,b)}^{logic} = g(f(\hat{I}_a),f(I_b))
\end{aligned}
\end{equation}

where $f(\cdot)$ denotes the semantic encoder, implemented using the mge-small model, and $g(\cdot)$ denotes the similarity function, computed via cosine similarity.In addition to the interest accuracy reward, we further introduce a format reward for the reasoning process, requiring the model to strictly follow the predefined three-step reasoning procedure specified in the instruction. Specifically, we detect the presence of the key tokens ["Step1", "Step2", "Step3"] and parse each step in JSON format to verify whether all required attribute keys are complete. If the output satisfies the format constraints, $R_{think\_format}=1$; otherwise, $R_{think\_format}=0$.

\subsubsection{Query-Level Reward}
The query-level reward is designed to supervise the generation of guidance queries that are both commercially valuable and content-compliant. To this end, we design two types of rewards and apply them to each generated guidance query $q^k_j$ under every sub-interest $I_k$.The first is a length reward $R_{length}$. We require the guidance query to contain no more than 15 Chinese characters to ensure concise expression. If the requirement is satisfied, $R_{length}=1$; otherwise, $R_{length}=0$.

The second is the shopping-guidance value reward $R_{value}$. We first train a judge model on manually annotated data to evaluate the commercial guidance value of each guidance query. The judge model assigns a rating $rt \in \{S,A,B\}$, and different rewards are assigned according to the corresponding rating level.

\begin{equation}
\begin{aligned}
R_{value} = \begin{cases}
1, & \text{if } rt =S \\
0.5, & \text{if } rt = A \\
0,  & \text{if } rt = B
\end{cases}
\end{aligned}
\end{equation}

The criteria for each rating level are defined as follows:

\fbox{
\parbox{0.4\textwidth}{
\textbf{S Level:} Targeting specific consumer groups or usage scenarios; Reflecting decisions; Helping users understand differences in item parameters comparison. \\
\textbf{A Level:} Contain flagship store channel words; Advertising statement sentence; The content is vague and general.\\
\textbf{B Level:} The sentence is not smooth; Non e-commerce content; The answer cannot be mapped to a real item; No information difference; Brand category mismatch; Content involving politics, pornography, and other unsafe elements.  
}
}

\subsubsection{Interest-Level Reward}
The interest-level reward is designed to supervise the content accuracy and stylistic diversity of multiple candidate guidance queries within the same interest cluster. To this end, we design three reward functions.

The first is the semantic accuracy reward $R_{sem}$. Assume that the reference guidance queries generated by the high-parameter model under sub-interest $I_k$ is denoted as $\hat{q}_k$, while the model generates $T$ candidate guidance queries $q_k=[q_{k_1},q_{k_2},...,q_{k_T}]$ for the same interest. Meanwhile, Step 3 provides the confidence scores $\{c_k\}_{i=1}^T$ for each guidance query. We first normalize the confidence scores to obtain the importance weights $\{w_k\}_{i=1}^T$ for each guidance query, and then compute the semantic similarity between each candidate guidance query $q_{k_j}$ and the reference guidance query $\hat{q}_k$. The final semantic accuracy reward $R_{sem}$ is obtained through weighted aggregation of these similarity scores.

\begin{equation}
\begin{aligned}
R_{sem} = \sum_{i=1}^Tw_{k_i}g(f(q_{k_i}), f(\hat{q}_k)), \omega_{k_i}=\frac{c_{k_i}}{\sum_{j=1}^Tc_{k_j}}
\end{aligned}
\end{equation}

where $k$ denotes the corresponding interest cluster, $g(\cdot)$ represents the similarity function implemented using cosine similarity, and $f(\cdot)$ denotes the semantic encoder, implemented with the mge-small model.

In addition, we design a stylistic diversity reward $R_{style}$ to encourage varied expressions of guidance queries within the same interest cluster. We adopt the Type-Token Ratio (TTR) to measure stylistic diversity, where the ratio of unique tokens to the total number of tokens in $q_k=[q_{k_1},q_{k_2},...,q_{k_T}]$ is used as the value of $R_{style}$.

\begin{equation}
\begin{aligned}
R_{style} = \frac{|\cup_{i=1}^T\{Token(q_{k_i})\}|}{\sum_{i=1}^T|Token(q_{k_i})|}
\end{aligned}
\end{equation}

Finally, we introduce a simple format constraint reward $R_{format}$, which requires the number of candidate guidance queries $N_q$ generated for each interest cluster to exactly match the instructed quantity $T$, preventing the model from generating too few or too many guidance queries.

\begin{equation}
\begin{aligned}
R_{format} = \begin{cases}
1, & \text{if } N_q = T \\
0, & \text{otherwise} 
\end{cases}
\end{aligned}
\end{equation}

\subsubsection{Global-Level Reward}
The global-level reward is designed to supervise content diversity across different interests. Specifically, we aim to encourage diversified interests and avoid homogeneous expressions among interest clusters. To this end, we introduce a content diversity reward $R_{content}$.

Assume that the model outputs $N$ interest groups $I_u=[I_1,I_2,...,I_N]$, where each sub-interest $I_j$ contains $T$ guidance queries $Q_j=[q_1,q_2,...,q_T]$. We first apply mean pooling over the semantic representations of the guidance queries within each group to obtain the semantic representation of the corresponding interest cluster:$E(\bar{Q}_j)=MeanPool(E(q_1),E(q_2),...,E(q_T))$.We then compute the semantic similarity between different interest clusters to obtain the similarity matrix $E_Q \in R^{N\times N}$. Finally, the value of $R_{content}$ is calculated based on $E_Q$. The corresponding formulas are defined as follows:

\begin{equation}
\begin{aligned}
R_{content} = 1 - \frac{1}{N(N-1)}\sum_{i=1}^N\sum_{j=1,j\neq i}^NE_{ij}
\end{aligned}
\end{equation}

where $E_{ij}$ denotes the semantic similarity between $E(\bar{Q}_i)$ and $E(\bar{Q}_j)$.

\subsubsection{Multi-Reward based GRPO Optimization}
We first integrate the multiple rewards described above using weighted aggregation, and then optimize the model with the GRPO algorithm. The formulation of the response-level reward is given below.

\begin{equation}
\begin{aligned}
&R_{response} = R_{think} + R_{query} + R_{interest} + R_{global} \\
&R_{think} = \omega_{think\_format}R_{think\_format} + \omega_{match}R_{match} \\
&R_{query} = \frac{1}{N*T}\sum_{i=1}^{N*T}(\omega_{length}R_{length}^i +\omega_{value}R_{value}^i) \\
&R_{interest} = \frac{1}{N}\sum_{i=1}^N(\omega_{sem}R_{sem}^i + 
\omega_{style}R_{style}^i + \omega_{format}R_{format}^i) \\
&R_{global} = \omega_{content}R_{content}
\end{aligned}
\end{equation}

where $N$ denotes the number of interest clusters inferred by the model, and $T$ denotes the number of guidance queries generated under each interest cluster.$\omega_*$ denotes the weight of the corresponding reward function. The objective function for this stage is defined as follows:

\begin{equation}
\begin{aligned}
& \mathcal{J}_{grpo} = E[\frac{1}{G}\sum_{i=1}^Gmin(r_i(\theta)A_i,
clip(r_i(\theta),1-\epsilon, 1+\epsilon)A_i)-\beta D_{KL}] \\
& r_i(\theta)=\frac{\pi_\theta(o_i|q)}{\pi_{old}(o_i|q)}, A_i=\frac{R_i-mean(\{R_1,R_2,...,R_G\})}{std(\{R_1,R_2,...,R_G\})+\epsilon}
\end{aligned}
\end{equation}

where $o_i$ denotes the i-th sampled output sequence, $q$ represents the model input, $\pi_\theta$ denotes the current policy, and $\pi_{old}$ denotes the previous policy. 

\subsection{DPO Training: Distill the Reasoning Ability of Think Mode}
After reinforcement learning with multi-reward supervision, the model achieves substantial improvements in both interest segmentation accuracy and the quality of generated guidance queries. However, in online environments where real-time user interactions require high responsiveness, multi-step reasoning during inference can significantly increase latency. Therefore, in practical deployment, we expect the model to directly generate the final answers without explicit reasoning steps.

Experimental results show that when deep reasoning is removed, the quality of the generated guidance queries deteriorates noticeably. This phenomenon can be attributed to the conditional distribution shift caused by the discrepancy between training and inference output formats, which consequently leads to degraded inference performance.

To address this issue, we introduce DPO training to distill the model’s reasoning capability acquired under the thinking mode through preference alignment. Specifically, We first use the model checkpoint from RL stage generate responses under the thinking mode and retain only their final answers as positive samples $y_w$, while discarding the explicit reasoning traces. Responses directly generated under the non-thinking mode are used as negative samples $y_l$. To enhance discriminability, 70\% of the negative samples are collected from checkpoints of the RL stage, while the remaining 30\% are sampled from the base model. After constructing the positive and negative sample pairs, the optimization objective of the model is defined as follows:

\begin{equation}
\begin{aligned}
\mathcal{J}_{DPO}(\pi_\theta;\pi_{ref})=-E[log \sigma(\beta log\frac{\pi_\theta(y_w|x)}{\pi_{ref}(y_w|x)}-\beta log\frac{\pi_\theta(y_l|x)}{\pi_{ref}(y_l|x)})]
\end{aligned}
\end{equation}

where $\pi_\theta$ denotes the policy model, $\pi_{ref}$ denotes the reference model, and $\beta$ is the temperature coefficient used to control the degree of deviation from the reference model.

During training, we observed that the preference margin between positive and negative samples was not sufficiently significant. To further enhance discriminability, we applied a simple preference-pair filtering strategy. Specifically, we compute the semantic similarity for each positive-negative sample pair and remove the top 30\% most semantically similar pairs to alleviate the issue of 'ineffective training'. Offline experimental results demonstrate that this preference-pair filtering strategy can improve the effectiveness of DPO training.

\subsection{Online Deployment}
To satisfy the high-throughput and low-latency requirements of online environments, we design both online and nearline serving pipelines to balance accuracy and responsiveness. The nearline pipeline is responsible for dynamically generating candidate guidance queries using the LLM4AIGQ framework and storing them in the mapping table, while the online pipeline handles real-time request processing through lightweight matching and ranking mechanisms to achieve fast response times.

\subsubsection{Nearline Recall}
The nearline pipeline is not responsible for real-time guidance queries generation. Instead, the model is invoked after every $N$ accumulated user interactions, such as clicks or favorites, to generate guidance queries. The generated guidance queries are directly written back into the mapping table for subsequent online retrieval. The model operates in non-thinking mode, directly outputting multiple interest labels along with their corresponding guidance queries. Meanwhile, the user historical behavior sequence is continuously updated as new interactions are triggered.

\subsubsection{Online Recall}
The online pipeline is responsible for real-time query serving. We construct an offline mapping table from user\_id to guidance queries, which dynamically receives the generated results from the nearline pipeline. Whenever a user enters the shopping-guide interface, the online pipeline employs an efficient vector retrieval algorithm to locate the corresponding index in the table and retrieve personalized candidate guidance queries for real-time recall.
\section{Experiment}
In this section, we conduct extensive offline experiments and rigorous online A/B tests on real-world industrial datasets to validate the effectiveness of LLM4AIGQ.

\subsection{Experiment Settings}
\subsubsection{Dataset}
Our datasets are sampled from multi-behavior user interaction logs on the Taobao and Tmall e-commerce platforms, covering the period from June to July 2026. The training data for the short-title fine-tuning task is collected from the Taobao item CPV table, with 8,000 sampled instances. For SFT training, we sample 5,000 single-day interaction records from highly active users. To support single-interest fine-tuning, each sample is decomposed into multiple single-interest samples using a strong teacher model (DeepSeek-V4-Pro~\cite{xu2026deepseek}), resulting in 11,259 training instances in total. For RL training, 3,000 samples are collected. For DPO training, we collect 5,000 samples, yielding 3,500 valid samples after preference-pair filtering. In addition, 1,000 test samples are collected for offline evaluation. To ensure data isolation, interaction data for different stages are sampled from different time partitions.

\subsubsection{Evaluation Metrics}
Given the lack of universal evaluation metrics for AI-generated  guidance queries, we mainly evaluate two aspects: relevance and shopping guidance value. For shopping guidance value, we report the proportions of S/A/B ratings for the generated queries on the test set. For relevance, we expect the generated queries to accurately match items that users are likely to interact with in the future. Specifically, for each test sample, we collect and mask the most recent 10 interacted items as positive samples $P_i$, and randomly sample 500 items from the whole item pool as negative samples $N_i$. For each generated query $q_i$, semantic similarities are computed against all samples $(P_i, N_i)$. We then examine whether the positive samples $P_i$ appear in the top-k ranked results. Recall@k and NDCG@k are adopted to measure the relevance of the generated queries.

In addition, considering that user interests may be complex and diverse, the sampled positive items may exhibit substantial interest discrepancies from the historical behaviors. To improve evaluation reliability, we compute a semantic similarity matrix $S_{hp}$ between the interaction history $H_i$ and the positive samples $P_i$, and use the average pooling result of $S_{hp}$ as a confidence score $c$. Assume there are $R$ test samples, each associated with $T$ generated queries, with $L$ history behaviors, with $M$ positive samples and $N$ negative samples. The formula for confidence score $c_i$ is defined as follows:

\begin{equation}
\begin{aligned}
c_i = \frac{1}{L_iM}\sum_{a=1}^{L_i}\sum_{b=1}^MS_{hp}^{(i)}(a,b)
\end{aligned}
\end{equation}

The formula for Recall@k is defined as follows:

\begin{equation}
\begin{aligned}
Recall@k = \frac{\sum_{i=1}^Rc_i\sum_{t=1}^T\sum_{j=1}^M1(rank^t_i(j) \leq k)}{\sum_{i=1}^Rc_i\cdot T \cdot M}
\end{aligned}
\end{equation}

Here, $rank^t_i(j)$ denotes the ranking position of the positive sample $p_i^j$ in the retrieved results, and $1(\cdot)$ is the indicator function, which equals 1 if the condition holds and 0 otherwise. The formula for NDCG@k is defined as follows:

\begin{equation}
\begin{aligned}
&NDCG@k = \frac{\sum_{i=1}^Rc_i\sum_{t=1}^T\frac{DCG_i^t@k}{IDCG@k}}{\sum_{i=1}^Rc_i \cdot T} \\
&DCG_i^t@k = \sum_{j=1}^M\frac{1(rank^t_i(j) \leq k)}{log2(rank^t_i(j) + 1)} \\
&IDCG@k = \sum_{j=1}^{min(M,k)}\frac{1}{log_2(j+1)}
\end{aligned}
\end{equation}

\begin{table*}[htbp]
\centering
\caption{Performance comparison among Different Models on Taobao Dataset with Relation and Value Metrics}
\label{table1}
\setlength{\aboverulesep}{0pt}
\setlength{\belowrulesep}{0pt}
\renewcommand{\arraystretch}{1.25} 

\begin{tabular}{l | cccccc | ccc}
\toprule
\multirow{2}{*}{Model} & \multicolumn{6}{c|}{Relation Metric} & \multicolumn{3}{c}{Value Metric} \\
\cmidrule(lr){2-7} \cmidrule(lr){8-10}
 & Recall@10 & NDCG@10 & Recall@20 & NDCG@20 & Recall@50 & NDCG@50 & S ratio & A ratio & B ratio \\
\midrule
Qwen3-30B-A3B & 0.1513 & 0.1754 & 0.1982 & 0.1980 & 0.2909 & 0.2331 & 0.6405 & 0.1043 & 0.2551 \\
Qwen3.5-122B-A10B & 0.1366 & 0.1580 & 0.1807 & 0.1792 & 0.2748 & 0.2146 & 0.8251 & 0.0401 & 0.1348 \\
Qwen3-235B-A22B & 0.1412 & 0.1641 & 0.1872 & 0.1862 & 0.2826 & 0.2222 & 0.7940 & 0.0461 & 0.1599 \\
GPT-oss-120B & 0.1227 & 0.1420 & 0.1658 & 0.1627 & 0.2563 & 0.1967 & 0.7177 & 0.0729 & 0.2093 \\
Deepseek-V4-Pro & 0.1363 & 0.1582 & 0.1805 & 0.1794 & 0.2708 & 0.2135 & 0.8152 & 0.0718 & 0.1130 \\
\midrule
LLM4AIGQ-$\textit{MULTI}_{\textit{SFT}}$ & 0.1674 & 0.1914 & 0.2183 & 0.2159 & 0.3125 & 0.2515 & 0.8044 & 0.0559 & 0.1396 \\
LLM4AIGQ-$\textit{ONE}_{\textit{SFT}}$ & 0.1709 & 0.1956 & 0.2189 & 0.2187 & 0.3127 & 0.2541 & 0.8119 & 0.0562 & 0.1318 \\
LLM4AIGQ-$\textit{MULTI}_{\textit{SFT+RL}}$ & 0.1719 & 0.1963 & 0.2221 & 0.2204 & 0.3179 & 0.2567 & 0.8138 & 0.0555 & 0.1307 \\
LLM4AIGQ-$\textit{ONE}_{\textit{SFT+RL}}$ & 0.1771 & 0.2034 & 0.2256 & 0.2267 & 0.3201 & 0.2625 & 0.8360 & 0.0520 & 0.1119 \\
LLM4AIGQ-$\textit{MULTI}_{\textit{SFT+RL+DPO}}$ & 0.2075 & 0.2364 & 0.2637 & 0.2635 & 0.3604 & 0.3002 & 0.8555 & 0.0296 & 0.1149 \\
LLM4AIGQ-$\textit{ONE}_{\textit{SFT+RL+DPO}}$ & \textbf{0.2101} & \textbf{0.2377} & \textbf{0.2684} & \textbf{0.2658} & \textbf{0.3709} & \textbf{0.3047} & \textbf{0.8616} & 0.0274 & 0.1109 \\
\bottomrule
\end{tabular}
\end{table*}

\subsubsection{Baseline Methods}
We compare LLM4AIGQ with two major types of baselines: (1) zero-shot LLMs, including several advanced models and larger variants from the same model family, all evaluated without any domain adaptation; and (2) variants of LLM4AIGQ. Specifically, we report the performance of model checkpoints from different training stages based on the single-interest SFT branch (ONE) and the multi-interest SFT branch (MULTI).

\subsubsection{Implementation Details}
LLM4AIGQ adopts Qwen3-30B-A3B as the backbone model and is trained on NVIDIA H20 GPUs. The maximum input and output lengths are set to 8192 and 2048, respectively. During SFT training, the model is trained for 500 steps using 4 H20 GPUs, with a per-device batch size of 2, gradient accumulation steps of 8, LoRA~\cite{hu2021lora} rank of 16, and learning rate of 1e-5. During the reinforcement learning stage, we use the ROLL framework~\cite{wang2025reinforcement} training for 100 steps on 16 H20 GPUs, with a group size of 8, learning rate of 1e-6, and vLLM~\cite{kwon2023efficient} as the inference engine. DPO training is conducted with LoRA fine-tuning for 500 steps using 4 H20 GPUs and gradient accumulation steps of 4. The reward weights in RL training are configured as follows: $w_{think_format}$ is 0.5, $w_{match}$ is 2.0, $w_{length}$ is 0.2, $w_{value}$ is 0.8, $w_{sem}$ is 0.5, $w_{style}$ is 0.3, $w_{format}$ is 0.2, and $w_{content}$ is 0.5.All offline experiments are conducted in non-think mode using the same instruction settings and the same testset.

\subsection{Model Performance}
As shown in Table ~\ref{table1}, we compare various baseline models and different variants of our method on the Taobao test set, from which the following conclusions can be drawn:
\begin{itemize}[leftmargin=*] 
\item Our method outperforms zero-shot state-of-the-art models in both relevance and shopping guidance value, and also surpasses larger models from the same family, demonstrating the effectiveness of LLM4AIGQ. The three-stage training paradigm of “SFT + RL + DPO” significantly improves interest mining and shopping guidance quality assessment. Notably, larger state-of-the-art models show worse relevance performance than the 30B backbone model, possibly because they tend to generate more novel queries, while smaller models rely more on high-frequency contextual terms. In e-commerce scenarios, such novelty is more prone to hallucination, which is harmful to shopping guidance tasks.

\item Comparisons among LLM4AIGQ variants show that instruction tuning on single-interest sequences consistently outperforms tuning on mixed-interest sequences, and this advantage remains throughout later training stages. We attribute this to the higher noise and information dilution in mixed-interest scenarios, where the model spends more effort identifying dominant interests instead of learning high-quality shopping guidance query patterns.

\item Each training stage significantly improves model performance. In the SFT stage, style learning rapidly enhances the value and relevance of generated AIGQ. In the RL stage, multi-level reward functions help the model disentangle user interests and improve relevance, while a judge model further enhances query value by supervising quality. The DPO stage improves reasoning in non-thinking mode by distilling reasoning ability from think mode, indirectly confirming weaker reasoning performance under non-thinking settings.

\end{itemize}

\begin{table}[htbp]
\centering
\caption{Analysis of the Response Performance of Item Short Titles to LLM4ALGQ}
\label{table3}
\begin{tabular}{lcccc}
\toprule
Ablation  & Input Tokens & Output Tokens & TTFT & RT \\
\midrule
short\_title & 2.35k & 146 & 110ms & 2.84s \\
long\_title & 3.21k & 153 & 114ms & 3.18s \\
\bottomrule
\end{tabular}
\end{table}

\begin{table}[htbp]
\centering
\caption{The Experimental Results of Ablation Study about GRPO-Training and DPO-Training.}
\label{table2}
\begin{tabular}{lccc}
\toprule
Model & Recall@10 & NDCG@10 & S ratio \\
\midrule
LLM4AIGQ-$\textit{ONE}_{\textit{SFT+RL}}$ & 0.1771 & 0.2034 & 0.8360 \\
w\_o\_think\_reward & 0.1696 & 0.1942 & 0.8332 \\
w\_o\_query\_reward & 0.1727 & 0.1980 & 0.8168 \\
w\_o\_interest\_reward & 0.1713 & 0.1972 & 0.8207 \\
w\_o\_global\_reward & 0.1744 & 0.2001 & 0.8195 \\
\midrule
LLM4AIGQ-$\textit{ONE}_{\textit{SFT+RL+DPO}}$ & 0.2101 & 0.2377 & 0.8616 \\
w\_o\_pair\_filter & 0.1959 & 0.2182 & 0.8426 \\
LLM4AIGQ-$\textit{MULTI}_{\textit{SFT+RL+DPO}}$ & 0.2075 & 0.2364 & 0.8555 \\
w\_o\_pair\_filter & 0.1922 & 0.2127 & 0.8297 \\
\bottomrule
\end{tabular}
\end{table}

\subsection{Ablation Study}


\subsubsection{The Effectiveness of Context Compression}
We extracted 100 samples from test set and used item short titles and original titles to construct user context. We conducted stress tests online and collected the response of the model. From Table ~\ref{table3}, it can be seen that when using short titles, the average length of the input token in the model will be reduced by 26.8\%, the response delay (RT) will be reduced by 10.7\%, and the model inference will be significantly faster. And the output token quantity will also slightly decrease. After case analysis, we found that the length and style of the model's inference of the guidance query will also imitate the item title. Therefore, when the item title becomes shorter, the model output will be more concise.

\subsubsection{The Effectiveness of Multi-Level Rewards in GRPO}
As shown in Table ~\ref{table2}, removing any reward component leads to declines in both relevance and shopping guidance value. In particular, removing the think-level reward causes a significant drop in relevance metrics, indicating that supervising interest disentanglement during the reasoning process helps the model more accurately capture users’ multiple interests and generate queries better aligned with user intent. Removing the query-level reward significantly degrades the value metrics, showing that the length constraint and the judge model’s quality assessment effectively guide the model toward generating more compliant and higher-quality shopping guidance queries. When the interest-level reward is removed, both relevance and value metrics decrease. This is because the absence of semantic correctness constraints weakens the model’s relevance judgment, while the lack of diversity guidance causes generated queries within the same interest cluster to become overly homogeneous in form, thereby reducing their value. Removing the global-level reward makes queries from different interests semantically similar to each other, which further harms the overall shopping guidance value.

\subsubsection{The Effectiveness of Data Filtering in DPO}
As shown in Table ~\ref{table2}, removing the preference-pair filtering strategy in DPO training leads to significant declines in both relevance and shopping guidance value. This is because the unfiltered positive-negative sample pairs contain many semantically similar cases. Training on such hard-to-distinguish data makes the optimization process more unstable and causes the model to produce more templated and stylistically homogeneous outputs, resulting in “ineffective learning” and ultimately weakening the effectiveness of distillation.

\begin{table}[htbp]
\centering
\caption{Pairwise Evaluation Result between LLMAIGQ variants and Teacher Model. The experiment adopts A/B randomization to prevent positional bias in the experimental results.}
\label{table4}
\begin{tabular}{lcc}
\toprule
model & Win Rate & Lose Rate \\
\midrule
base model & 34.63\% & 65.37\% \\
$LLM4AIGQ_{SFT}$ & 50.25\% & 49.75\% \\
$LLM4AIGQ_{SFT+RL}$ & 57.53\% & 42.47\% \\
$LLM4AIGQ_{SFT+RL+DPO}$ & 54.95\% & 45.05\% \\
\bottomrule
\end{tabular}
\end{table}

\subsection{A/B Pairwise Evaluation}
To further evaluate the overall quality of generated AIGQ beyond retrieval-oriented automatic metrics, we conduct an LLM-as-Judge pairwise evaluation by Gemini-3.5-Flash. Given the same user context, the judge compares the AIGQ generated by each LLM4AIGQ variant against those generated by the teacher model(deepseek-V4-Pro), considering their consistency with users' consumption intent and preferences, shopping-guidance value, content rationality, and linguistic quality.

As shown in Table ~\ref{table4}, the base model achieves only a 34.63\% win rate against the teacher model, indicating a large gap in domain-specific AIGQ generation. After SFT, the win rate rises to 50.25\%, showing that supervised learning effectively transfers the teacher’s domain knowledge and generation patterns to the student model. After RL, the win rate further increases to 57.53\%, allowing the student model to surpass the teacher. This suggests that the performance gain comes not only from imitating the teacher, but also from optimizing interest understanding, query quality, and content diversity.
After DPO, the win rate slightly drops from 57.53\% to 54.95\%, but still remains above the teacher model. This is consistent with the goal of DPO, which focuses on transferring reasoning capability from RL to efficient non-thinking inference rather than further improving generation quality. The small decline reflects a trade-off between reasoning quality and inference efficiency, while most RL-acquired capability is preserved.


\subsection{Online A/B Test}
We conducted online A/B tests across multiple business scenarios in Tmall, including shopping guidance, search, and item detail pages. Experimental results show that introducing AIGQ consistently improves user click performance in the query recommendation pipeline. In the initial 10\% traffic experiment, the experimental group achieved a 4.46\% improvement in uCTR compared with the baseline bucket. After scaling the traffic to 40\%, the overall uCTR still maintained a positive gain of 2.53\% over six days of online validation, demonstrating strong generalization capability and online stability under larger-scale deployment.


\section{Conclusion}

In this work, to address issues in traditional shopping guidance query generation pipelines, such as semantic drift and the inability to capture personalized user intent, we propose an LLM-based AI guidance query generation framework for multi-interest users. The framework can autonomously disentangle user interests and generate high-quality shopping guidance queries tailored to users’ consumption intentions, and capabilities. Our approach has already been deployed online and has achieved significant results at the scale of hundreds of millions of users.


\bibliographystyle{ACM-Reference-Format}
\bibliography{reference}

\appendix

\end{document}